\documentclass[10pt,superscriptaddress,nofootinbib,notitlepage,twocolumn,pra]{revtex4-2}

\usepackage{graphicx}
\usepackage{subfigure}
\usepackage{braket}
\usepackage{amssymb}
\usepackage{amsmath}
\usepackage{bbm}
\usepackage[colorlinks,anchorcolor=black,citecolor=blue,linkcolor=red,urlcolor=blue]{hyperref}

\begin{document}

\title{Discrete-Modulated Continuous-Variable Quantum Key Distribution in Satellite-to-Ground Communication}

\author{Shi-Gen Li}
\author{Chen-Long Li}
\author{Wen-Bo Liu}
\affiliation{National Laboratory of Solid State Microstructures and School of Physics, Collaborative Innovation Center of Advanced Microstructures, Nanjing University, Nanjing 210093, China}
\affiliation{Department of Physics and Beijing Key Laboratory of Opto-electronic Functional Materials and Micro-nano Devices, Key Laboratory of Quantum State Construction and Manipulation (Ministry of Education), Renmin University of China, Beijing 100872, China}
\author{Hua-Lei Yin}\email{hlyin@ruc.edu.cn}
\affiliation{Department of Physics and Beijing Key Laboratory of Opto-electronic Functional Materials and Micro-nano Devices, Key Laboratory of Quantum State Construction and Manipulation (Ministry of Education), Renmin University of China, Beijing 100872, China}
\affiliation{National Laboratory of Solid State Microstructures and School of Physics, Collaborative Innovation Center of Advanced Microstructures, Nanjing University, Nanjing 210093, China}
\author{Zeng-Bing Chen}\email{zbchen@nju.edu.cn}
\affiliation{National Laboratory of Solid State Microstructures and School of Physics, Collaborative Innovation Center of Advanced Microstructures, Nanjing University, Nanjing 210093, China}
\date{\today}

\begin{abstract}
Satellite-to-ground quantum communication constitutes the cornerstone of the global quantum network, heralding the advent of the future of quantum information. Continuous-variable quantum key distribution is a strong candidate for space-ground quantum communication due to its simplicity, stability, and ease of implementation, especially for the robustness of space background light noise. Recently, the discrete-modulated continuous-variable protocol has garnered increased attention, owing to its lower implementation requirements, acceptable security key rate, and pronounced compatibility with extant infrastructures. Here, we derive key rates for discrete-modulated continuous-variable quantum key distribution protocols in free-space channel environments across various conditions through numerical simulation, revealing the viability of its application in satellite-to-ground communication.
\end{abstract}

\maketitle

\section{Introduction}
The global quantum network, interconnecting quantum devices through quantum channels, possesses significant advantages in terms of security~\cite{wehner2018quantum,scarani2009security,yin2023experimental,lu2022micius,cao2024experimental,de2002quantum,weng2023beating,zhang2017quantum}. Similar the transformative impact of the internet on the global landscape, the quantum network will bring a renovation to the world. Numerous applications support this potential, including confidential communication and digital signatures~\cite{fitzsimons2017private,wang2022twin,sheng2022one,zhou2023experimental,liu2023experimental,li2023one,roehsner2018quantum}. Distributed quantum computation is feasible under its inherent power and distant quantum sensors can be connected~\cite{yehia2023connecting,denchev2008distributed}. Satellite-to-ground communication~\cite{bonato2009feasibility,moli2009performance,meyer2011implement,bourgoin2013comprehensive,boone2015entanglement,hosseinidehaj2018satellite,bedington2017progress,nauerth2013air,wang2013direct,bourgoin2015experimental,vallone2015experimental} is a crucial component in the establishment of a quantum network. In contrast to optical fiber channels, the satellite-to-ground channel offers a distinct advantage in long-distance communication with photon signal, as its transmissivity decreases at a more gradual rate~\cite{pirandola2021limits}. This characteristic contributes to its efficacy in facilitating secure quantum communication over extended distances. Many efforts have been put into the research of practical experiment~\cite{villoresi2008experimental,yin2013experimental,dequal2016experimental,carrasco2016leo,takenaka2017satellite,calderaro2018towards}, though we still have a long way to go till concrete applications get into something real.

Quantum key distribution (QKD), ensuring communication security, has drawn enormous attention since it was proposed~\cite{comandar2016quantum,haseler2010quantum,yin2020experimental,yin2016measurement}. QKD protocols can be broadly categorized into discrete variable (DV) and continuous variable (CV) classes~\cite{zhang2024continuous}. DV QKD protocols have made great achievements in both experimental and theoretical domains. CV QKD, however, has garnered considerable attention from researchers due to its compatibility with modern optical facilities, regardless of temporary disadvantages in communication distance. Notably, the requirement for single-photon encounters is alleviated. Protocols with squeezed states which are discrete~\cite{hillery2000quantum} and Gaussian~\cite{cerf2001quantum} encoded are firstly proposed, followed by Gaussian-modulated protocols with coherent states~\cite{weedbrook2012gaussian,grosshans2002continuous,grosshans2003quantum}, which avoid the challenge of production of squeezed light. Significantly, research in CV QKD employing Gaussian modulation has made remarkable advancements in experimental aspects~\cite{zhang2020long,lodewyck2007quantum,fossier2009field,jouguet2013experimental,huang2015high,qi2015generating,soh2015self,huang2016long,kleis2017continuous,zhang2019continuous,laudenbach2019pilot,wang2018high,wang2018long,liu2020continuous}. Moreover, with well-established security proofs~\cite{pirandola2015high,diamanti2015distributing,zhang2020one} and the development of finite-size analysis methods~\cite{guo2018channel,kish2020feasibility,villasenor2020atmospheric}, there have been an undeniable progress in theoretical aspects. Nonetheless, this kind of protocol encounters challenges due to its substantial demands on modulation devices and relatively low efficiency in error correction~\cite{lin2019asymptotic}. Then, discrete-modulated CV QKD is proposed with many efforts on security proofs~\cite{zhao2009asymptotic,bradler2018security,fan2023quantum}. Compared with Gaussian modulation protocols, discrete-modulated protocols mitigate the demand for extensive modulation devices. Hence, its appropriateness for application in satellite-based communication is underscored, particularly given the satellite's sensitivity to equipment weight.

\begin{figure*}[ht!]
    \centering
    \includegraphics[width=0.8\linewidth]{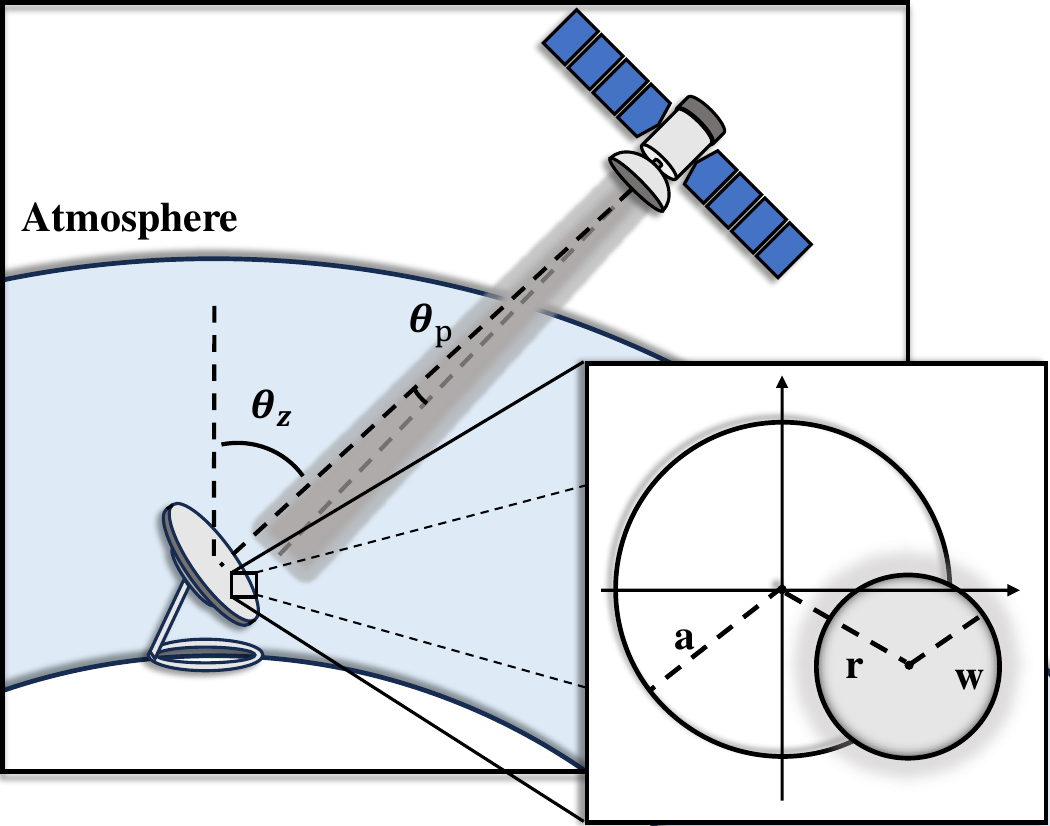}
    \caption{A sketch of Satellite-to-ground downlink channel. The satellite emits the laser beam with a point error $\theta_p$. Then the beam arrives at the ground station after passing through the outer space and the atmosphere with a zenith angle $\theta_z$. Finally, the laser beam reaches the aperture $a$ with a beam waist of width $w$ and a deviation from the center denoted by $r$.}
    \label{fig:STG_model}
\end{figure*}
While discrete-modulated protocols offer several advantages, the achievable key rate tends to be slightly lower compared to Gaussian-modulated protocols. The purpose of this paper is to investigate the viability of discrete-modulated CV QKD protocols within the context of satellite-to-ground communication channels. The protocols we use entail quaternary modulation schemes, which are also called quadrature phase-shift keying schemes with homodyne detection~\cite{liu2021homodyne} and heterodyne detection~\cite{lin2019asymptotic}, respectively. In particular, this kind of homodyne protocol has been shown to exhibit superior tolerance to excess noise~\cite{liu2021homodyne}. We utilize the secret key rate to describe the performance of protocols and conduct an investigation into the factors that significantly influence the key rate. The influence of the satellite-to-ground channel on key rate can be analyzed distinctively in both downlink and uplink scenarios, determining the degree to which atmospheric conditions affect signal transmission.  In the uplink channel, the laser signal originating from the ground station traverses outer space after encountering initial atmospheric effects, whereas this sequence is reversed in the downlink channel (Figure~\ref{fig:STG_model}). As a result, atmospheric conditions exert a notably greater impact on the uplink signal. We analyze these atmospheric effects and their implications on efficiency and excess noise. We compose the performance of the discrete-modulated CV-QKD system in different protocol parameters and channel conditions. Finally, we compare key rates and discuss the viability of satellite-based communication with discrete-modulated CV QKD.

\section{Satellite-to-ground Channel}
In this context, we conduct an analysis of the atmospheric impacts on laser signals, which encompass attenuation, disturbance, and deflection of the signal. Within our simulation framework, the receiver receives the anticipated photon considering the transmittance denoted as $\eta$, along with an environmental mode represented by $\hat{e}$. The impact of environmental mode occurrences, leading to deviations in measurement outcomes, is captured through excess noise
\begin{equation}
    \xi = \frac{(\Delta q_{\rm obs})^2}{(\Delta q_{\rm vac})^2}-1,
\end{equation}
where $(\Delta q_{\rm obs})^2$ and $(\Delta q_{\rm vac})^2$ represent the variance of the $q$ quadrature of observed state and vacuum state.

Firstly, we examine the phenomenon of photon absorption within the atmosphere, which can be derived using the Beer-Lambert equation \cite{vasylyev2019satellite}
\begin{equation}
    \eta_{\rm ext}(h,\theta) = \exp\left\{-\int_{0}^{z} \alpha[h(y,\theta)] \,dy\right\},
\end{equation}
where $\alpha(h) = \alpha_0 e^{-h/\tilde{h}}$ represents the extinction function, with $h$ indicating the satellite's altitude. Considering both aerosol and molecular scattering, the extinction coefficient at sea level, denoted as $\alpha_0$, is computed as $4\times10^{-7}~\rm m^{-1}$ at a wavelength of $\lambda = 1550~\rm nm$ and $5\times10^{-6}~\rm m^{-1}$ at $\lambda = 800~\rm nm$. The scale parameter $\tilde{h}$ is specified as $6600~\rm m$.

Furthermore, the atmosphere introduces additional effects on the light beam due to its uneven distribution in both time and space, leading to phenomena such as beam broadening and beam wandering~\cite{gunthner2017quantum,vasylyev2019satellite}. Specifically, in the context of short-term exposure, we can observe the broadened beam waist $W_{\rm ST}$ and the shifted beam center caused by spatial turbulence \cite{dios2004scintillation}. Yura provides a useful approximation for the broadened beam waist in the uplink channel \cite{yura1973short} that is 
\begin{equation}
    \begin{split}
    W_{\rm ST}^2(z) &= W_0^2 \left(1+\frac{z^2}{Z_0^2}\right) + \\
    &\frac{35.28z^2}{k^2 r_s^2} \left[1-0.26 \left(\frac{r_s}{W_0}\right)^{1/3}\right]^2,
    \end{split}
\end{equation}
where $W_0$ represents the primary beam waist, $z$ is the propagation distance, $Z_0= kW_0^2/2$ denotes the Rayleigh distance, $k$ is the wavenumber, and $r_s$ is a factor representing turbulence intensity. The turbulence intensity factor $r_s$ can be determined as follows
\begin{equation}
    r_s = \left[ 0.42k^2\int_0^L C_n^2(h(z,\theta_z))\left(\frac{L-z}{L}\right)^{5/3} dz\right]^{-3/5},
\end{equation}
where $L$ denotes the total propagation distance and $C_n^2(h)$ represents the refractive index structure parameter. This expression is valid under the condition $0.26(r_s/W_0)^{1/3}<<1$. The refractive index structure parameter $C_n^2(h)$ can be modeled using the Hufnagel-Valley (HV) model \cite{andrews2005laser}, which is described by
\begin{equation}
\begin{split}
    C_n^2(h) &= 8.1481\times10^{-56} v^2 h^{10} e^{-h/1000} + \\
    &2.7\times10^{-16} e^{-h/1500} + C_0 e^{-h/100},  
\end{split}
\end{equation}
where $v$ is the wind speed, $h$ represents altitude, and $C_0=9.6\times 10^{-14} \rm m^{-2/3}$~\cite{qing2016use} is a constant term. 
\begin{figure}[t]
    \centering
    \includegraphics[width=\linewidth]{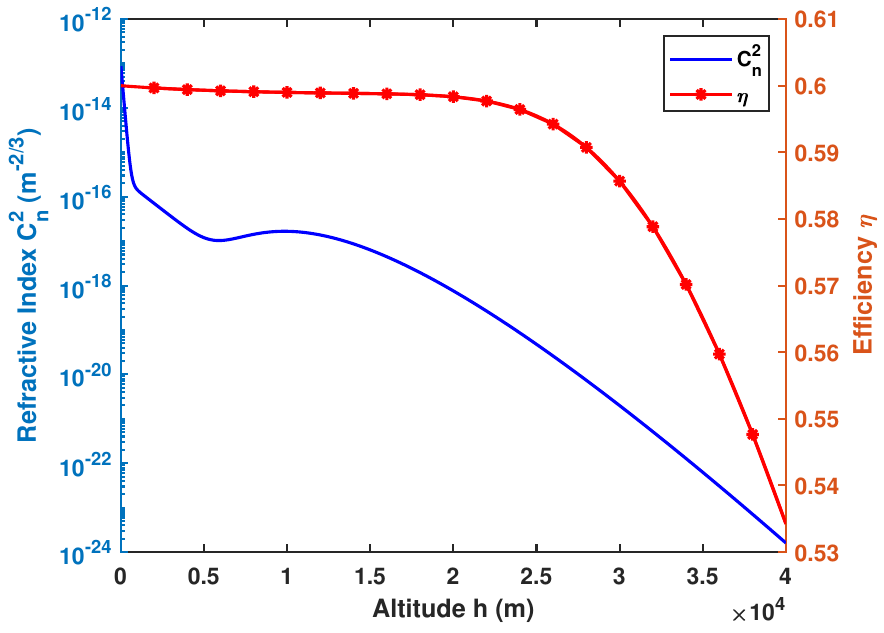}
    \caption{Plot of refractive index $C^2_n(h)$ near the earth's surface and the efficiency of uplink channel with parameters $\theta_z=0,\theta_p = 1 ~{\rm \mu rad}$~\cite{liao2017satellite}$,v_{\rm wind}=21~{\rm m/s}, a = 0.75~{\rm m}, W_0 = 0.2~{\rm m}$.}
    \label{fig:efficiency_uplink}
\end{figure}
Under long-term exposure, due to the unevenness in time, the deviation of the center $r=\sqrt{x^2+y^2}$ follows a random distribution given by
\begin{equation}
    P(x,y)= \frac{1}{2\pi \sigma_r^2} \exp\left(-\frac{x^2+y^2}{2\sigma_r^2} \right),
\end{equation}
where the variance $\sigma_r$ is a result of atmospheric effects and pointing error, defined as $\sigma_r^2 = (L\theta_p)^2+\sigma_{\rm TB}^2$. Here, $\theta_p$ represents the pointing error angle, and the variance $\sigma_{\rm TB}$ caused by atmospheric effects can be computed as
\begin{equation}
\sigma_{\rm TB}^2 = 1.035\int_0^L C^2_n(z)(L-z)^2 W_{\rm ST}^{-1/3}(z) dz.
\end{equation}
The efficiency $\eta(r)$ corresponding to each point $r$ is formulated as follows \cite{vasylyev2012toward}
\begin{equation}
\begin{split}
\eta(r) = &\frac{2}{\pi W^2}\exp\left(-2\frac{r^2}{W^2}\right) \\
&\cdot \int_0^a \rho\exp\left(-2\frac{\rho^2}{W^2}\right)I_0\left(\frac{4r\rho}{W^2}\right) d\rho,
\end{split}
\end{equation}
where $W$ represents the beam waist, specifically $W_{ST}(z=L)$, and $a$ denotes the radius of the receiver's aperture. The function $I_n(x)$ represents the n-th order modified Bessel function. Indeed, directly evaluating the integral in the efficiency expression can be challenging. There is an excellent approximation for this integral provided by Vasylyev's work~\cite{vasylyev2012toward}. By combining the probability distribution of deviation $r$ and the corresponding efficiency $\eta(r)$, we can readily compute the overall efficiency affected by atmospheric turbulence. This analysis is specifically tailored for the uplink channel, where the beam initially encounters significant influences, thus making atmospheric effects highly impactful on the received signal. The total efficiencies computed for typically used 800 nm and 1550 nm lasers, based on the provided parameters, are 0.0021 and 0.003, respectively. Given the widespread adoption of the 1550 nm laser in fiber communication due to its advantageous properties, we also opt to utilize it in the context of satellite-based communication. 

In the context of the downlink channel, factors such as pointing error and the consequent expansion of the beam waist play a significant role \cite{dequal2021feasibility}. The variance of the downlink channel can be estimated as $\sigma_{\rm TB}^2 \approx 1.919C_n^2z^3(2W_0)^{-1/3}$ \cite{dequal2021feasibility}. From Figure~\ref{fig:efficiency_uplink} in the paper, the refractive index typically falls within the range of $10^{-14} \sim 10^{-16} \, \mathrm{m^{-2/3}}$. Consequently, we observe that $\sigma_{\rm TB}^2$ is much smaller compared to $(L\theta_p)^2$. Since the atmosphere puts the strength in the last short distance, i.e., $W_{\rm ST}^2(z) \approx W_0^2(1+\frac{z^2}{Z_0^2})$~\cite{pirandola2021satellite}.

\begin{figure*}[ht]
    \centering
    \includegraphics[width=\linewidth]{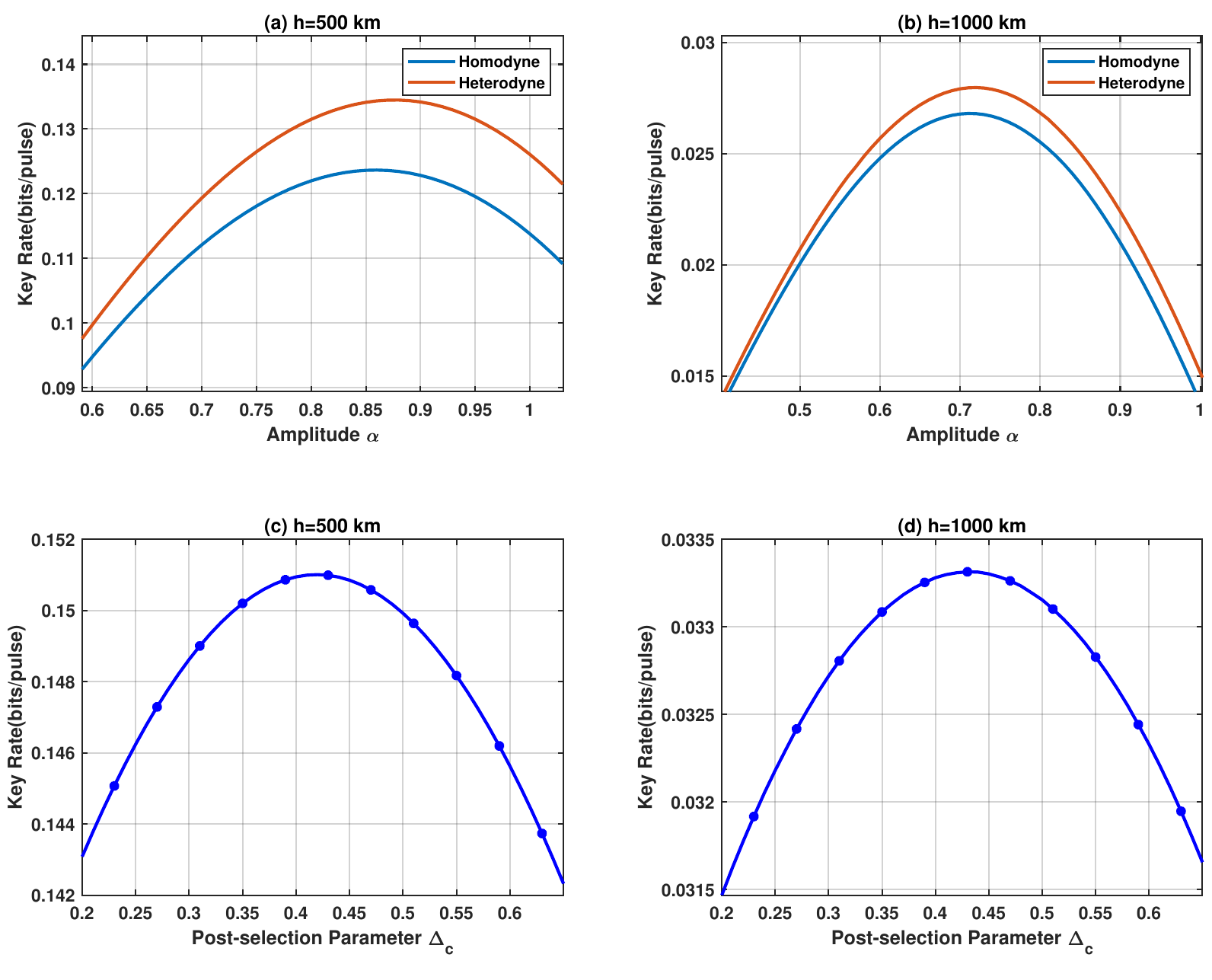}
    \caption{Asymptotic key rate variation versus protocol parameters under some typical conditions \textcolor{black}{with ideal detectors}. (a)(b) The asymptotic key rate variations are analyzed concerning the amplitude parameter $\alpha$ for both homodyne and heterodyne protocols at altitudes of 500 km and 1000 km in the downlink channel. (c)(d) The key rate variations are studied concerning the post-selection parameter $\Delta_c$ for the homodyne protocol with optimal $\alpha$ values at various satellite altitudes in the downlink channel.}
    \label{fig:alpha_delta_hom_optimization}
\end{figure*}
Apart from the factors above, imperfections in devices introduce photon loss during the processes of photon creation and the journey from the reception of photons to the generation of electrical signals. These defects arise due to the constraints of current technology. But these are not easily affected by the atmosphere, so we merge them and set them to a fixed value $\eta_{\rm dev}=0.6$~\cite{jovanovic2017efficient,pirandola2021limits,pirandola2021satellite}. Combining all the parts mentioned above, we could make an approximation of the efficiency of the total system.

\begin{figure*}[ht]
    \centering
    \includegraphics[width=\linewidth]{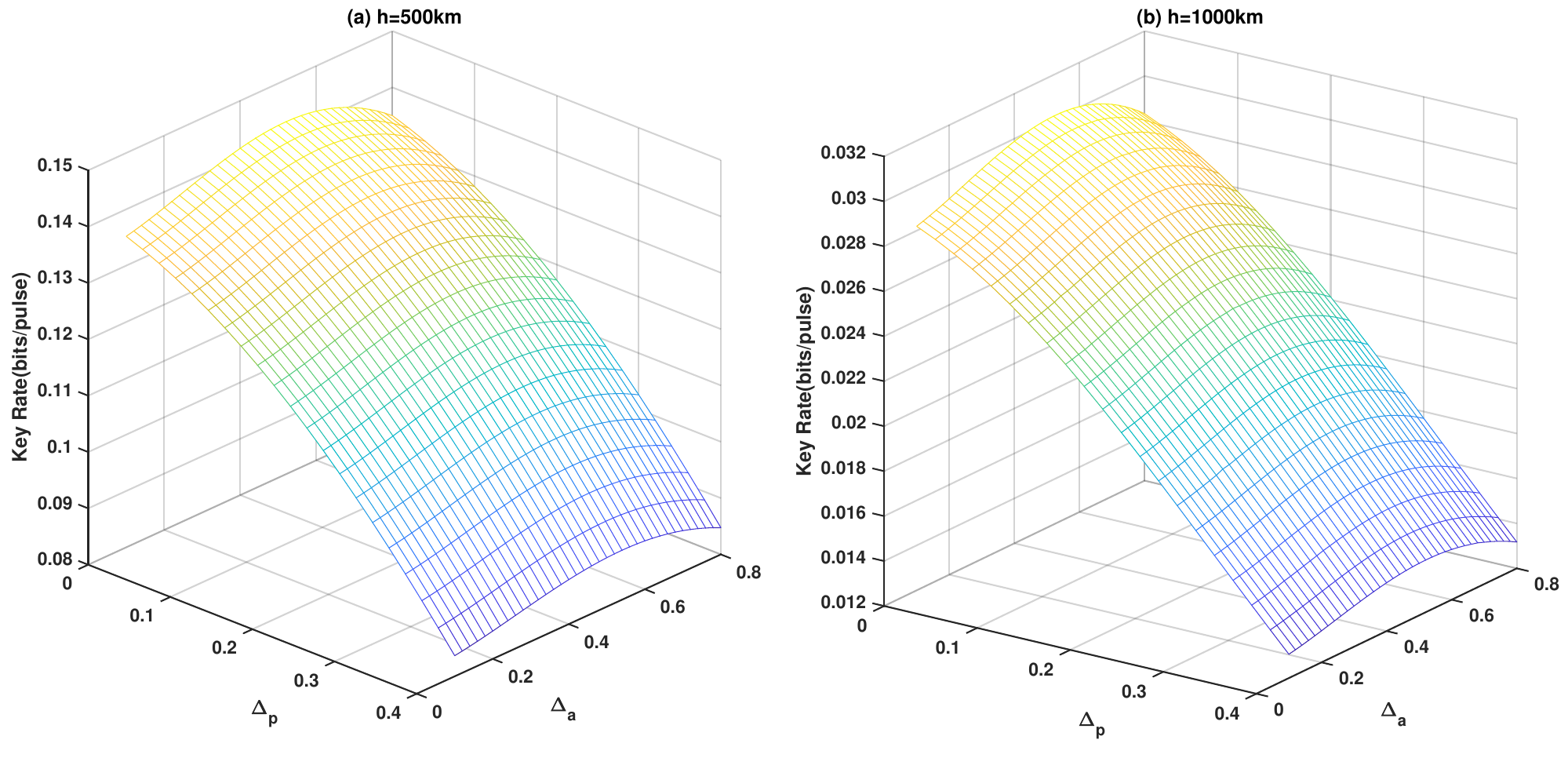}
    \caption{The secure key rate is evaluated with respect to the post-selection parameters of the heterodyne protocol, with the amplitude set to its optimal value. This simulation is conducted under the conditions of the downlink channel at an altitude of 500 km.}
    \label{fig:delta_het_optimization}
\end{figure*}
Next, we turn our attention to the excess noise in the satellite-to-ground channel. In discrete-modulated CV QKD protocols, excess noise originates from various sources and can be broadly classified into two groups. The first group consists of channel noise, which encompasses scintillation induced by atmospheric fluctuations and additional photons from sunlight or other light sources. The noise from the channel can fluctuate with changes in climate or variations in day and night conditions. This noise can be approximated by the temporarily received power of other light sources. Kish has provided typical noise intensities for us, as documented in their work \cite{kish2020feasibility}. The second group involves detection noise, where imperfections in detection devices lead to unforeseen errors, with certain types, such as electronic noise, being inherently unavoidable. The total excess noise can be expressed as $\xi = \xi_{\rm ch} + \xi_{\rm det}/\eta$~\cite{laudenbach2018continuous}. It is important to note that, due to the specific implementation of the detection scheme, the detection noise associated with heterodyne detection is twice that of homodyne detection. \textcolor{black}{As depicted in the equation, a decrease in efficiency leads to a significant increase in excess noise, consequently exerting a substantial negative influence on the secret key rate. Therefore, we also take into account the performance of ideal detectors. The secret key rate exhibited by ideal detectors closely resembles that of trusted detectors~\cite{lin2020trusted}. Trusted detectors, characterized by inefficiency and electronic noise, do not compromise information security to eavesdroppers, unlike untrusted detectors which, besides being imperfect, are susceptible to eavesdropping. Given that users typically have control over the detectors, this assumption closely mirrors real-world scenarios.}

\section{Discrete-modulated Protocols}
Here we make a brief description of the protocols. The protocols utilized in this investigation involve two discrete-modulated CV QKD protocols, employing both homodyne and heterodyne detection~\cite{liu2021homodyne,lin2019asymptotic}. In each iteration of the two protocols, Alice randomly prepares a state chosen from the set of coherent states \(\{\ket{\alpha e^{i\pi/4}},\ket{\alpha e^{3i\pi/4}},\ket{\alpha e^{5i\pi/4}},\ket{\alpha e^{7i\pi/4}}\}\) and transmits it to Bob. Upon receiving the state, Bob performs either heterodyne detection, employing a positive operator-valued measure \(\{E_{\gamma} = (1/\pi)\ket{\gamma}\bra{\gamma}:\gamma \in \mathcal{C}\}\), or homodyne detection. In the latter case, he randomly selects a quadrature from \(\{\hat{q},\hat{p}\}\) to measure. For homodyne detection outcomes, values less than $-\Delta_c$ are denoted as 0, and those greater than $\Delta_c$ are denoted as 1, while other values are discarded. In heterodyne detection, measurement outcomes are of the form $\gamma e^{i\theta}$. Those satisfying $\gamma > \Delta_a$ and $\Delta_p \leq \theta < \frac{\pi}{2}-\Delta_p$ are labeled as 0, while others are marked as 1, 2, 3 based on their quadrant. Subsequently, after reverse reconciliation and privacy amplification procedures, the parties obtain the secret keys.

\begin{figure*}[t]
    \centering
    \includegraphics[width=\linewidth]{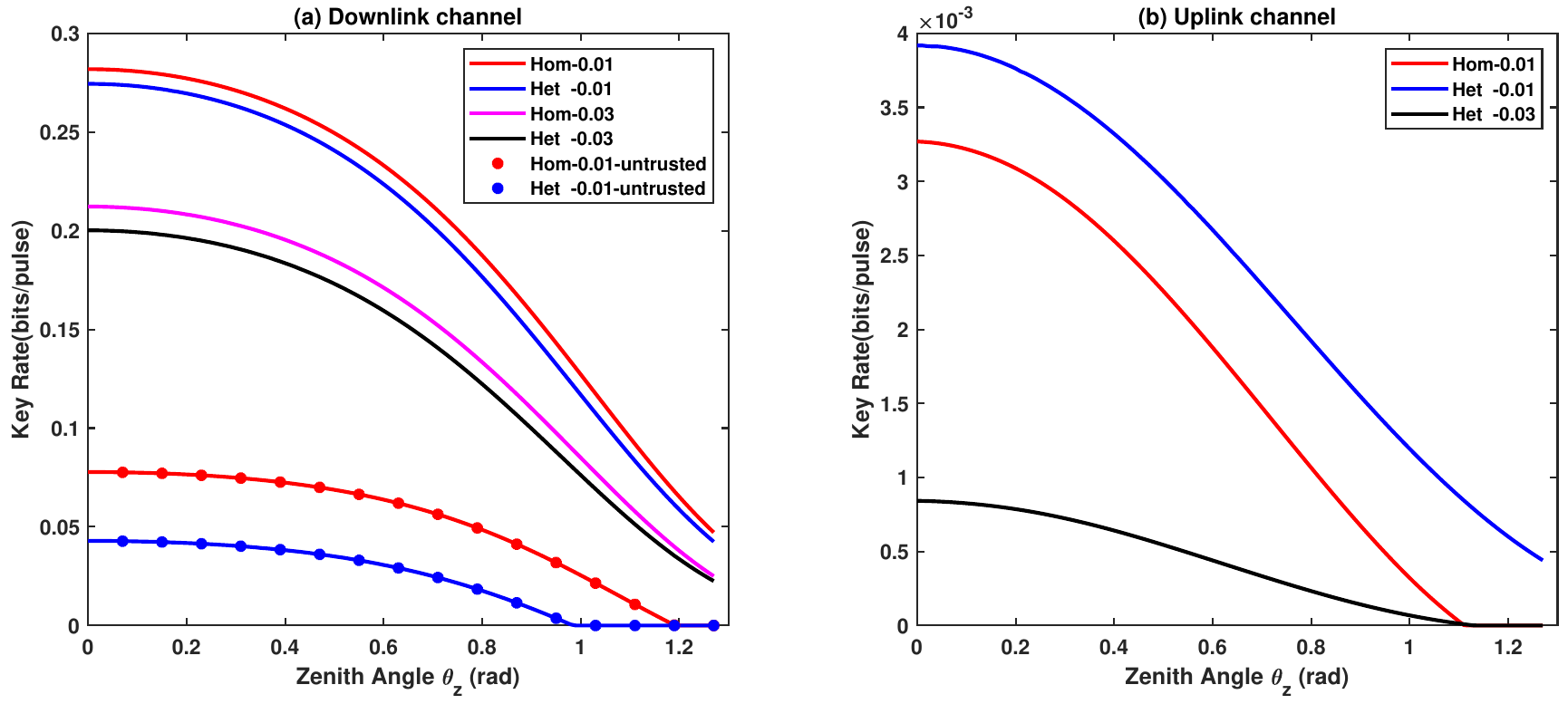}
    \caption{Secret key rate versus zenith angle $\theta_z$. Both plots are generated through simulations conducted at an altitude of 300 km. Plot (a) delineates the relationship between the key rate and zenith angle, considering an excess noise parameter of \textcolor{black}{$\xi_{\rm ch} = 0.01~\text{and}~0.03$} in the downlink channel. \textcolor{black}{Ideal and untrusted} detectors are both considered. Plot (b) illustrates variations in key rate within the uplink channel specifically employing \textcolor{black}{ideal detectors}.}
    \label{fig:performance_zenith}
\end{figure*}
The secret key rate serves as a measure of the security of the communication channel. Through the application of the source-replacement scheme~\cite{curty2004entanglement,yin2020experimental}, the transmitted state is transformed into a bipartite state $\rho_{AB}$. The key rate of an equivalent entanglement-based scheme, subject to asymptotic conditions~\cite{coles2016numerical,winick2018reliable}, is obtained using the Devetak-Winter formula~\cite{devetak2005distillation,grosshans2003quantum}
\begin{equation}
    R^\infty = p_{\rm pass}[\min_{\rho_{AB}\in S} H(Z|E)-H(Z|X)].
    \label{eq:keyrate}
\end{equation}
Here, $p_{\rm pass}$ represents the sifting probability, which is the probability that a given round is retained. $H(Z|E)$ denotes the conditional von Neumann entropy of the raw key string $Z$ given Eve's knowledge, and $H(Z|X)$ is the conditional von Neumann entropy with $X$ representing Alice's string. The set of constraints, denoted by $S$, outlines the conditions under which the protocol operates. The solving of the secret key rate can be transformed into a convex optimization problem, which is described comprehensively in Appendix.

\section{Simulation Results}
Initially, we will make adjustments to the parameters utilized within the protocols. Compared with Gaussian-modulated protocols, discrete-modulated CV QKD protocols introduce additional post-selection parameters denoted as $\Delta_c$, $\Delta_a$, and $\Delta_p$. These parameters, alongside the amplitude $\alpha$, collectively exert a substantial influence on the secure key rate. We explore specific scenarios where the satellite is positioned at altitudes of 500 km and 1000 km. The satellite emits a beam with a primary beam waist of $W_0=0.2~{\rm m}$ and the previously mentioned wavelengths toward the receiver, whose aperture radius is 0.75 m located at sea level. The beam traverses the atmosphere, encountering excess noise characterized by $\xi_{\textcolor{black}{\rm ch}} = 0.01$, a pointing error of $10^{-6} \mu{\rm m}$, and a zenith angle of $\theta_z = 0$. Parameters such as the wind speed ($v=21{\rm m/s}$) and photon number cut-off ($N_c=10$) are set, defined by the state $\ket{\alpha}=\sum_{n=0}^{10} \frac{\alpha^n}{\sqrt{n!}}\ket{n}$. Subsequently, we generate the following figures based on the simulation outcomes, considering the efficiency of error correction $\beta=0.95$ due to advancements in technology. From the analysis depicted in Figure~\ref{fig:alpha_delta_hom_optimization}, we observe a trend where the secure key rate initially increases and then decreases as the amplitude varies. Notably, the peaks occur at \textcolor{black}{$\alpha = 0.86$ and $0.72$ for altitudes of 500 km and 1000 km, respectively.} Upon optimizing the amplitude, we determine that the optimal post-selection parameter $\Delta_c$ for the homodyne protocol is approximately within the range of [0.4, 0.45], as illustrated in Figure~\ref{fig:alpha_delta_hom_optimization}. It is important to note that the optimal values of $\alpha$ and $\Delta_c$ decrease with an increase in satellite altitude. For the upcoming simulations starting from 100 km, given that the key rate remains relatively stable around the optimal values, we directly set \textcolor{black}{$\alpha = 0.72$ and $\Delta_c=0.42$}. In the case of the heterodyne protocol, the post-selection parameters $\Delta_a$ and $\Delta_p$ are utilized to filter signals based on amplitude and phase. By searching for the maximal key rate with \textcolor{black}{$\Delta_a \in [0.1, 0.8]$ and $\Delta_p \in [0, 0.4]$} with a step size of 0.02, we observe from Figure~\ref{fig:delta_het_optimization} that the key rate decreases as $\Delta_p$ increases, reaching its maximum value at approximately \textcolor{black}{$\Delta_a=0.52$} and $\Delta_p=0$ for both scenarios. Therefore, we adopt these optimal values for the subsequent simulation configurations. 

\begin{figure*}[t]
    \centering
    \includegraphics[width=\linewidth]{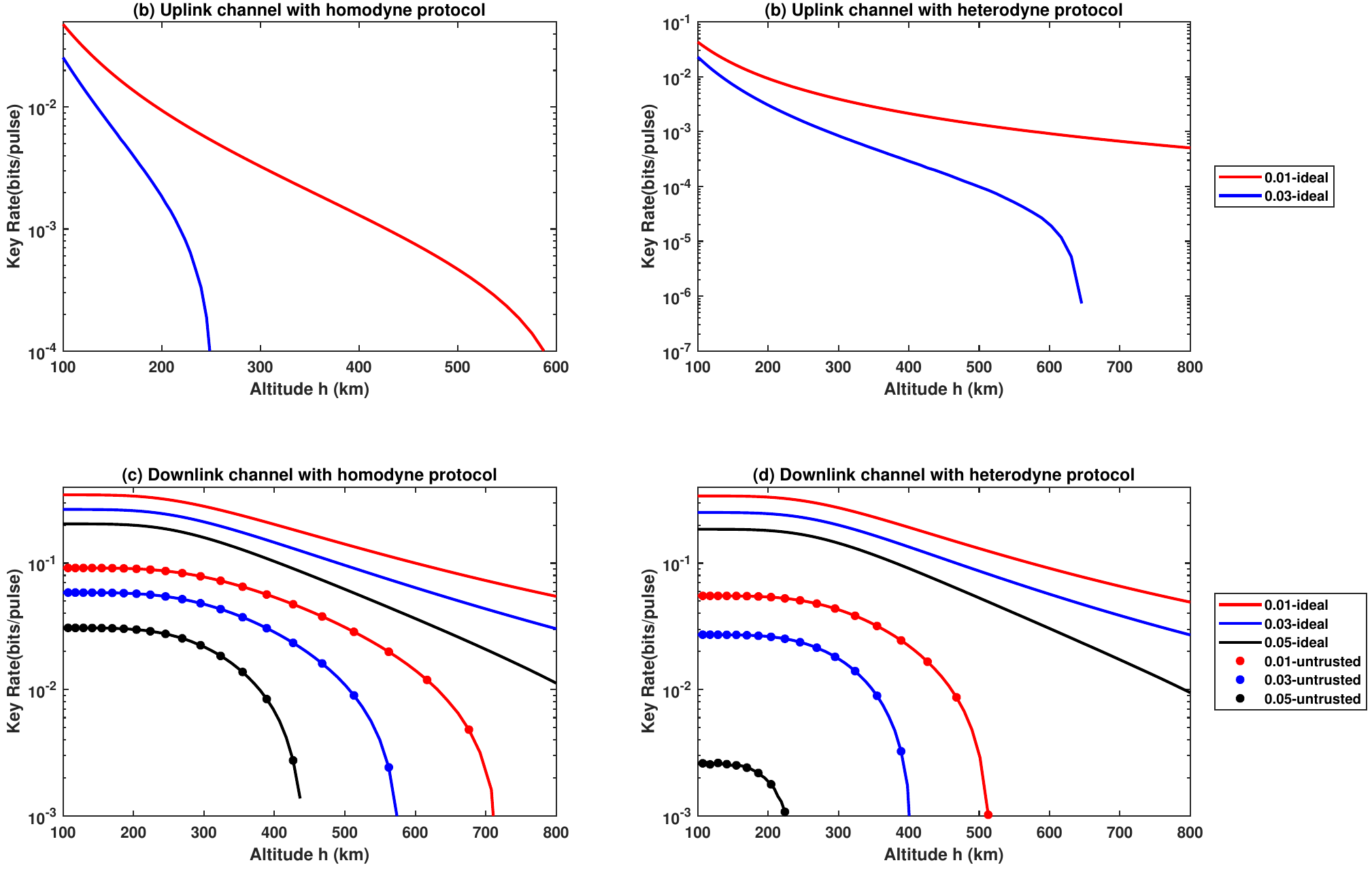}
    \caption{Secret key rate versus altitude with different channels and detection methods. Plots (a) and (b) collectively represent the key rates associated with distinct protocols within the uplink channel, employing \textcolor{black}{ideal detectors and considering varying levels of channel excess noise}. Conversely, plots (c) and (d) are dedicated to delineating the variability in key rate within the downlink channel, incorporating \textcolor{black}{channel}~excess noise levels of $0.01$, $0.03$, and $0.05$, and assessing both \textcolor{black}{ideal}~and untrusted detectors. All plots are generated under a consistent zenith angle of 0. The variation in excess noise settings across the plots represents different background noise conditions, contributing to a comprehensive evaluation of system performance under varying environmental factors.}
    \label{fig:performance_altitude}
\end{figure*}
We then proceed to examine the performance of discrete-modulated protocols under various conditions. Besides atmospheric and surface conditions, the satellite's position plays a crucial role. The zenith angle is particularly significant as it determines the duration of secret key generation communication, directly affecting the resulting key length. In Figure~\ref{fig:performance_zenith}, we present simulations depicting the variation of the secret key rate as a function of the zenith angle at an altitude of 300 km under the downlink channel. The protocols exhibit considerable resilience to fluctuations in the zenith angle. It is also evident that the secure key rate drops significantly as excess noise increases. The key rate of the protocol with homodyne detection is generally higher than that of the protocol with heterodyne detection \textcolor{black}{in condition with high effciency}. Scenarios where key rates are negative are not depicted in the figure.

In Figure~\ref{fig:performance_altitude}, we explore the performance under the downlink and uplink channels with different noise conditions. In the first two plots, the performance of the uplink channel is presented with homodyne and heterodyne protocols together with ideal detectors. In the last two plots, the performance of the downlink channel is revealed. We consider detectors' noise $\xi_{\rm het}=2\xi_{\rm hom}=2\xi_{\rm det}=0.02$ and channel noise $\xi_{\rm ch}=0.01,0.03,0.05$. The selection of noise values comes from Ref.~\cite{kish2020feasibility}. The simulation curves of the uplink channel are fewer than those of the downlink channel because the efficiency of the uplink channel is much lower than that of the downlink channel, resulting in a weak tolerance to excess noise. However, it is worth noting that the aperture receives significantly fewer extra photons in outer space, so the excess noise should be much lower in the uplink channel. The heterodyne protocol possesses a higher key rate at higher altitudes \textcolor{black}{with ideal detectors in uplink channel}, while it performs poorly when considering detector noise due to its detection pattern. Overall, as seen from the figures, the protocols exhibit an acceptable secure key rate under different conditions. 

\section{Discussion}
Discrete-modulated CV QKD has drawn great interest due to its low requirements on modulation devices leading to the potential of being applied in satellite-to-ground communication, which is an exciting piece of news for future applications and an immense potential for satellite-based communication considering the satellite's demanding requirements on weight. We aim to figure out its practical performance in possible channels. In the preceding sections, we provided a concise overview of the channel model and simulation method. And finally, we get the particular performance of discrete-modulated protocols in various scenarios. By manipulating parameters such as zenith angle, transmission direction, satellite altitude, excess noise intensity, and detector conditions, the protocols exhibit a positive key rate against various negative factors. The resilience to photon loss and excess noise affords the feasibility of placing the satellite in Low Earth Orbit (LEO) and establishing communication in intricate conditions. Compared with Gaussian-modulated protocols, discrete-modulated protocols can achieve comparable communication efficiency with lower demands on modulation devices.

In future research, theoretical investigations could focus more on understanding the finite-size effect of discrete-modulated protocols, as this area remains unexplored compared to the well-developed analysis of Gaussian-modulated protocols. Moreover, there is a need for more precise estimations regarding environmental effects, which can be achieved with the support of experimental data. In terms of practical implementation, there is a critical need for detectors with low noise levels to effectively manage the issue of excess noise.

\medskip
\section*{Appendix: Simulation Method of Secret Key Rate}

The Eq.~\ref{eq:keyrate} of heterodyne detection protocol can be reformulated as a convex optimization problem~\cite{coles2016numerical,winick2018reliable}
\begin{equation}
    R^\infty = \min_{\rho_{AB} \in S} D(\mathcal{G}(\rho_{AB})||\mathcal{Z}[\mathcal{G}(\rho_{AB})])-p_{\rm pass}\delta_{EC}.
\end{equation}
Here, $D(\rho||\sigma)=\rm Tr(\rho\log_2(\rho))-\rm Tr(\rho\log_2(\sigma))$ represents the quantum relative entropy, $\mathcal{G}(\rho)$ represents the post-processing procedure, and the error correction consumption $\delta_{EC}$ can be calculated from the equation $\delta_{EC} = H(Z)-\beta I(X;Z) = (1-\beta)H(Z) + \beta H(Z|X)$ with an error correction efficiency of $\beta=0.95$.

In the context of heterodyne detection, the expression for the map $\mathcal{G}(\rho)$ is given by $\mathcal{G}(\rho)=K\sigma K^\dag$, where the corresponding Kraus operator is delineated as
\begin{equation}
    K = \sum_{z=0}^3 \ket{z}_R \otimes \mathbbm{1}_A \otimes (\sqrt{R_z})_B,
\end{equation}
with
\begin{equation}
    R_z = \frac{1}{\pi} \int_{\Delta_a}^\infty d\gamma   \int_{z\pi/2+\Delta_p}^{(z+1)\pi/2-\Delta_p} \gamma\ket{\gamma e^{i\theta}}\bra{\gamma e^{i\theta}}d\theta.
\end{equation}
The provided equation implies that if the measurement outcome belongs to the image space corresponding to the projection operator $\sqrt{R_z}$, which indicates that the outcome has successfully passed through the post-selection process, the register state $R$ is modified to the corresponding state $\ket{z}_R$. For a more comprehensive discussion, further details can be found in references~\cite{winick2018reliable,lin2019asymptotic}. The parameters $\Delta_a$ and $\Delta_p$ denote the post-selection parameters. The function $\mathcal{Z}(\rho)$, named the pinching quantum channel, is defined by the projection operators $Z_j = \ket{j}\bra{j}\otimes \mathbbm{1}_{AB}$ as
\begin{equation}
    \mathcal{Z}(\rho) = \sum_{j=0}^{3} Z_j \rho Z_j.
\end{equation}
In the context of homodyne detection, the procedure differs somewhat. The raw key string is concurrently derived from the $p$ and $q$ quadratures. The key rate is expressed as
\begin{equation}
\begin{split}
    R^\infty = & \frac{1}{2} \left\{\min_{\rho_{AB} \in S} \sum_{y\in \{q,p\}} D(\mathcal{G}_y(\rho_{AB})||\mathcal{Z}[\mathcal{G}_y(\rho_{AB})]) \right.\\
    &\left.-\sum_{y\in \{q,p\}}p^y_{\rm pass}\delta_{\rm EC}^y\right\}.   
\end{split}
\end{equation}
The post-processing mapping $\mathcal{G}_y(\rho) = K_y \rho K_y^\dag$ is defined as
\begin{equation}
    K_y  = \sum_{b=0}^{1} \ket{b}_R \otimes \mathbbm{1}_A \otimes (\sqrt{I_y^b})_B .
\end{equation}
In this expression, $\{I_y^b\}$ denotes a set of projective operators
\begin{equation}
    I_y^0 = \int_{\Delta_c}^\infty dt\ket{t}\bra{t}, ~ I_y^1= \int_{-\infty}^{-\Delta_c} dt \ket{t} \bra{t}.
\end{equation}
It is noteworthy that $\ket{t}$ represents the eigenstate with eigenvalue $t$ of the quadrature $y$.
The constraints of the convex problem are specified as follows. First, the state should adhere to the observation results
\begin{equation}
\begin{split}
    &\textcolor{black}{{\rm Tr}[\rho_{AB}(\ket{x}\bra{x}_A\otimes \hat{q})] = p_x \braket{\hat{q}}_x,}\\
    &\textcolor{black}{{\rm Tr}[\rho_{AB}(\ket{x}\bra{x}_A\otimes \hat{p})] = p_x \braket{\hat{p}}_x,}\\
    &\textcolor{black}{{\rm Tr}[\rho_{AB}(\ket{x}\bra{x}_A\otimes \hat{n})] = p_x \braket{\hat{n}}_x,}\\
    &\textcolor{black}{{\rm Tr}[\rho_{AB}(\ket{x}\bra{x}_A\otimes \hat{d})] = p_x \braket{\hat{d}}_x.}\\
\end{split}
\end{equation}
Here, the operators $\hat{n}=\frac{1}{2}(\hat{q}^2+\hat{p}^2-1)$ and $\hat{d} = \hat{q}^2-\hat{p}^2$ are defined, with the relationships $\hat{q}$ and $\hat{p}$ being expressible in terms of creation and annihilation operators, simulated in Fock space under the photon-number cutoff assumption~\cite{lin2019asymptotic}. The mean values $\braket{\hat{q}}_x$, $\braket{\hat{p}}_x$, $\braket{\hat{n}}_x$, and $\braket{\hat{d}}_x$ are functions of the efficiency $\eta$ and excess noise $\xi$. For $\alpha_x \in \{\alpha e^{i\pi/4},\alpha e^{3i\pi/4},\alpha e^{5i\pi/4},\alpha e^{7i\pi/4}\}$, the following relations hold
\begin{equation}
\begin{split}
    &\braket{\hat{q}}_x = \sqrt{2\eta}\rm Re(\alpha_x),\\
    &\braket{\hat{p}}_x = \sqrt{2\eta}\rm Im(\alpha_x),\\
    &\braket{\hat{n}}_x = \eta|\alpha_x|^2+\frac{\eta\xi}{2},\\
    &\braket{\hat{d}}_x = \eta[\alpha_x^2+(\alpha_x^*)^2].    \label{equation_qpnd}
\end{split}
\end{equation}
The probability $p_x$ corresponds to the likelihood of state $\ket{x}$, and for this paper, it is assumed that $p_x=0.25$. Subsequently, as Alice prepares the quantum state and transmits half of it to Bob, the condition mandates that the state of Alice's subsystem attains an ideal configuration. Hence, the rule is expressed as follows
\begin{equation}
    \textcolor{black}{{\rm Tr}_B[\rho_{AB}]} = \sum_{i,j=0}^{3} \sqrt{p_i p_j} \braket{\phi_j|\phi_i}\ket{i}\bra{j}_A.
\end{equation}
Ultimately, $\rho_{AB}$ denotes the density matrix representing the composite state, and as such, it must satisfy the fundamental properties inherent to density matrices
\begin{equation}
\begin{split}
        &\textcolor{black}{{\rm Tr}[\rho_{AB}]} = 1,\\
        &\rho_{AB} \ge 0. 
\end{split}
\end{equation}
Moreover, the constraints can be reformulated as
\begin{equation}
    S = \{\rho \in \mathcal{H}_+ | \rm Tr(\Gamma_i\rho)=\gamma_i, ~\forall i\},
\end{equation}
where $\mathcal{H}_+$ represents the set of positive semi-definite operators.

The key rate needs to cut the error correction consumption. To compute this consumption $p_{\rm pass}^y\delta_{\rm EC}^y$, the conditional probability distribution of quadrature measurement outcomes $H(Z|X)$ is needed. Outcomes of homodyne detection is determined by
\begin{equation}
\begin{split}
    P(q|x) &= \frac{1}{\sqrt{\pi(\eta\xi+1)}}e^{[-(q-\sqrt{2\eta}\rm Re(\alpha_x))^2]/(\eta\xi+1)},\\
    P(p|x) &= \frac{1}{\sqrt{\pi(\eta\xi+1)}}e^{[-(p-\sqrt{2\eta}\rm Im(\alpha_x))^2]/(\eta\xi+1)}.
\end{split}
\end{equation}
The probability distribution of Bob's bits is expressed as
\begin{equation}
\begin{split}
        &P_y(0|x) = \int_{\textcolor{black}{\Delta_c}}^{\infty}P(y|x)dy,\\
        &P_y(1|x) = \int_{-\infty}^{\textcolor{black}{\Delta_c}}P(y|x)dy,\\
        &P_y(\perp|x) = \int_{-\textcolor{black}{\Delta_c}}^{\textcolor{black}{\Delta_c}}P(y|x)dy.
\end{split}
\end{equation}
Thus, by discarding bits labeled $\perp$, the sifting probability is given by
\begin{equation}
    p_{\rm pass}^y = \sum_{x}p_x(P_y(0|x)+P_y(1|x)),
\end{equation}
and the conditional entropy is expressed as
\begin{equation}
    H(Z_y|X_y) = \sum_{x}p_x h\left(\frac{P_y(0|x)}{P_y(0|x)+P_y(1|x)}\right),
\end{equation}
where $h(x)$ denotes the binary entropy.

For the heterodyne protocol, the error rate $p_{\rm pass}\delta_{\rm EC}$ can be evaluated based on the distribution
\begin{equation}
\begin{split}
         P(z&=j|x=k) = \textcolor{black}{{\rm Tr}(R_j\rho_B^k)}\\
        &= \int_{\Delta_a}^{\infty}d\gamma\int_{j\pi/2+\Delta_p}^{(j+1)\pi/2-\Delta_p}\frac{\exp\left(-\frac{|\gamma e^{i\theta}-\sqrt{\eta}\alpha_k|^2}{1+\eta\xi/2}\right)}{\pi\left(1+\eta\xi/2\right)}\gamma d\theta.
\end{split}
\end{equation}
By employing linear algebra and certain techniques~\cite{winick2018reliable,liu2021homodyne}, these convex optimization problems can be efficiently solved with CVX package~\cite{grant2014cvx,grant2008graph,toh1999sdpt3,tutuncu2003solving} in MATLAB.

\noindent {\bf Acknowledgments}\\

\noindent This work is supported by the National Natural Science Foundation of China (No. 12274223), the Program for Innovative Talents and Entrepreneurs in Jiangsu (No. JSSCRC2021484), and the Program of Song Shan Laboratory (included in the management of Major Science and Technology Program of Henan Province) (No. 221100210800-02).\\

\noindent{\bf Conflict of Interest}\\

\noindent The authors declare no conflict of interest.\\

\noindent{\bf Data Availability Statement}\\
The data that support the ﬁndings of this study are available from the corresponding author upon reasonable request.\\

\noindent{\bf Keywords}\\

\noindent Quantum key distribution, discrete-modulated, continuous-variable, satellite-to-ground, heterodyne detection


%

\end{document}